\documentclass{elsart}

\usepackage{epsfig,newlfont}

\usepackage{amssymb}
\usepackage{amsfonts,newlfont,epsfig}

\DeclareMathAlphabet{\mathitbf}{T1}{cmr}{bx}{it}
\def\bbox#1{{ \mbox{\boldmath $#1$}}}

\newcommand{\s}{{\mathrm{s}}}
\newcommand{\uu}{{\mathrm{u}}}
\newcommand{\cc}{{\mathrm{c}}}

\begin{document}
\begin{frontmatter}

\title{Dynamical generation of a gauge symmetry\\ in the Double-Exchange model}
\author[unizar]{J.M.~Carmona},
\author[unizar,inbifi]{A.~Cruz},
\author[ucm,inbifi]{L.~A.~Fern\'andez},
\author[unizar,inbifi]{S.~Jim\'enez},
\author[ucm,inbifi]{V. Mart\'{\i}n-Mayor},
\author[ucm]{A. Mu\~noz-Sudupe},
\author[unizar]{J.~Pech},
\author[unex,inbifi]{J.~J.~Ruiz-Lorenzo}
\author[unizar,inbifi]{A.~Taranc\'on}, and 
\author[unizarsia]{P.~T\'ellez}.

\address[unizar]{Departamento de F\'{\i}sica Te\'orica,
Facultad de Ciencias, \\
Universidad de Zaragoza, 50009 Zaragoza, SPAIN}
\address[ucm]{Departamento de F\'{\i}sica Te\'orica I, Facultad de CC. 
F\'{\i}sicas, \\
Universidad Complutense de Madrid, 28040 Madrid, SPAIN}
\address[unex]{Departamento de F\'{\i}sica, Facultad de Ciencias,\\
Universidad de Extremadura, 06071 Badajoz, SPAIN}
\address[unizarsia]{Servicio de Instrumentaci\'on Cient\'{\i}fica, 
Facultad de Ciencias,\\
Universidad de Zaragoza, 50009 Zaragoza, SPAIN}
\address[inbifi]{Instituto de Biocomputaci\'on y  F\'{\i}sica de
Sistemas Complejos (INBIFI),\\
Facultad de Ciencias, Universidad de Zaragoza, 50009 Zaragoza, SPAIN}

\date{27 December 2002}

\begin{abstract}
It is shown that a bosonic formulation of 
the double-exchange model, one of the classical
models for magnetism, generates dynamically a gauge-invariant phase in
a finite region of the phase diagram. We use analytical methods, Monte Carlo
simulations and Finite-Size Scaling analysis. We study the transition line
between that region and the paramagnetic phase.
The numerical results show that this transition line
belongs to the Universality Class of the Antiferromagnetic RP$^2$ model. 
The fact that one can define a
Universality Class for the Antiferromagnetic RP$^2$ model, different
from the one of the O($N$) models, is puzzling and somehow contradicts
naive expectations about Universality.
\end{abstract}

\end{frontmatter}

\section{Introduction}

The problem of the dynamical restoration of a gauge symmetry~(see
e.g.~\cite{GSR,PARISI} and references therein) has received
considerable attention in the recent 10 years, because of the problem
of introducing a chiral gauge theory in the lattice. Although the
Ginsparg-Wilson~\cite{GW} method has somehow superseded this approach, the
question remains an interesting one.  Naively, the problem could seem
a trivial one~\cite{PARISI}: the non gauge-invariant terms in the
action generate a high-temperature-like gauge-invariant expansion with
a finite radius of convergence. The subtle question is whether the
radius of convergence of this expansion will remain finite in the
continuum limit or not. In this letter, we want to address a related,
but different question, namely, the generation of a local invariance
in the low-temperature (broken-symmetry) phase of a system without any
explicitly gauge-invariant term in the action.  We have found this
intriguing phenomenon in a numerical study of a simplified version of
the Double-Exchange
Model~\cite{DEM,PLAN1}(DEM), one of the most general models for
magnetism in condensed matter physics, still under active
investigation~\cite{DAGOTTO}.  The local invariance does not follow
from a high-temperature-like expansion, but from the infinite
degeneracy of the ground-state (see next section), which occurs at a
unique value of the control parameter at zero temperature, then
extending to a finite region of the phase diagram at finite
temperature.  This phenomenon reminds one of the so-called
Quantum-Critical Point phenomenology~\cite{QCP}.

We have studied the model using Monte Carlo simulations and
Finite-Size Scaling techniques~\cite{LIBROFSS,ON,ISPERC}. We have
found that the critical exponents are fully compatible with the
ones~\cite{RP2D3} of the antiferromagnetic (AFM) RP$^2$ model in three
dimensions~\cite{RP2D3,ROMANO,SHROCK}, which has an explicit local
Z$_2$ invariance. This might not be surprising, given the strong
similarities in the ground-state of both models (see next
section). However, the fact that one can explicitly show that there is
a Universality Class associated to the AFM-RP$^2$ model is puzzling.
Indeed, the most ambitious formulation of the Universality Hypothesis
states that the critical properties of a system are fully determined
by the space dimensionality and its symmetry groups at
high-temperature ($\cal G$) and low-temperature ($\cal H$).  Moreover,
systems with a locally isomorphic $\cal G/\cal H$ are expected to have
the same critical behavior. In our case, $\cal G$=O(3), and from our
numerical study $\cal H$ seems to be O(2)~\cite{CRUZ}, although the
O(2) residual symmetry could be also broken (to O(1)=Z$_2$). In the
former case, the Universality Class should be the one of the O(3)
non-linear $\sigma$ model, while in the latter case one expects O(4)
non-linear $\sigma$ model-like behavior~\cite{AZARIA}. Our results are
definitely incompatible with an O(3)/O(2) scheme of symmetry breaking,
and very hardly compatible with an O(4)/O(3) (locally isomorphic to O(3)/O(1))
scheme.

In recent years, the Universality Hypothesis (as stated above) has
been challenged in a number of frustrated, chiral
models~\cite{KAWAMURA}. Yet, detailed numerical simulations have shown
that typical transitions are weakly first-order~\cite{ITAKURA}, which
is hardly surprising, because the typical critical exponents proposed
for chiral systems~\cite{KAWAMURA} are fairly similar to the
effective exponents one expects in weak first-order
transitions~\cite{WEAKFIRSTORDER}. On the other hand, we have no doubt
that the transition here studied is continuous, but we have no alternative
explanation for our results.

\section{The Model}

\begin{figure}[t]
\begin{center}
{\centering\epsfig{file=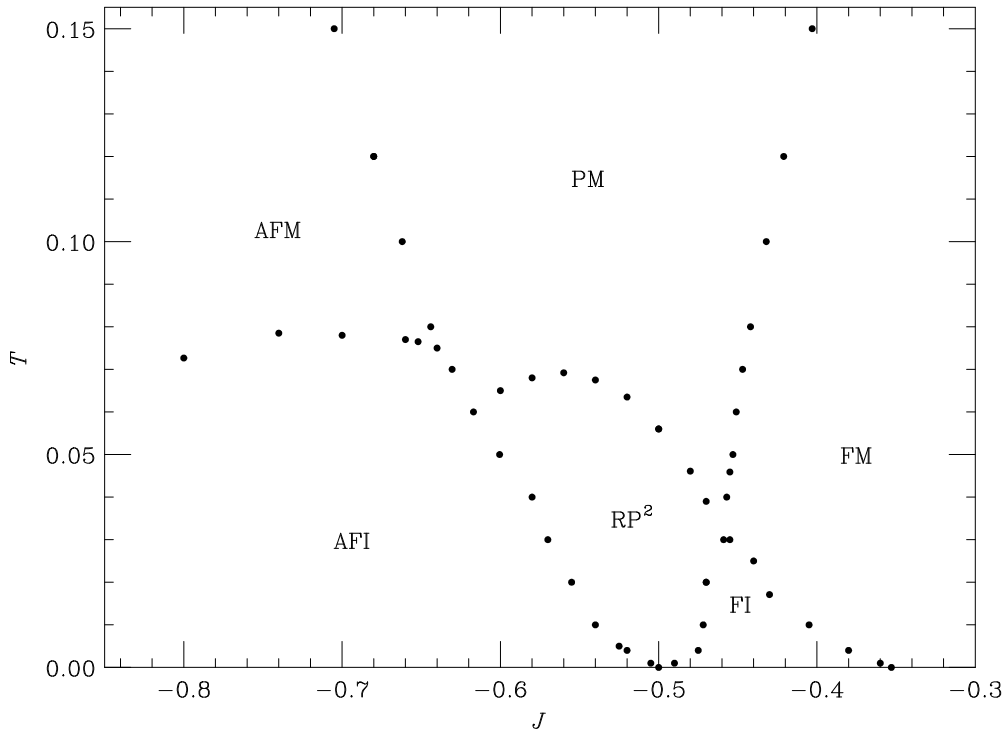,width=0.7\linewidth,angle=90}}
\end{center}
\protect\caption{Phase diagram of the model (\protect\ref{accion}), as
obtained from Monte Carlo simulation. The labels correspond to
paramagnetic (PM), ferromagnetic (FM), ferrimagnetic (FI),
antiferrimagnetic (AFI), antiferromagnetic (AFM), and RP$^2$-symmetric
(RP$^2$) phases.}
\label{FIGPHASEDIAG}
\end{figure}

Although some powerful techniques have been developed~\cite{OURDEM}
for the Double-Exchange model~\cite{DEM} (involving dynamical
fermions), lattice sizes beyond $L=16$ are extremely hard to study with 
the present generations of computers. Thus, one may turn to the
simplified version proposed by Anderson~\cite{DEM}, where a purely
bosonic Hamiltonian is considered. This simplified model has been
recently studied~\cite{PLAN1} by extensive Monte Carlo simulation. Yet,
previous studies missed several phases in the phase diagram (see
Fig.~\ref{FIGPHASEDIAG} and below). 

Specifically, we consider a three dimensional cubic lattice of side
$L$, with periodic boundary conditions. The dynamical variables,
$\vec\phi_i$, live on the sites of the lattice and are
three-component vectors of unit modulus. The Hamiltonian contains the
Anderson version of the Double-Exchange model plus an
antiferromagnetic first-neighbor Heisenberg
interaction~\cite{FENDEM}:
\begin{equation}
-H=\sum_{<i,j>}J\vec\phi_i\cdot \vec\phi_j +
\sqrt{1+\vec\phi_i\cdot \vec\phi_j}\,\quad,\quad J<0\,,
\label{accion}
\end{equation}
where $<\!i,j\!>$ means first-neighbor sites on the lattice. The partition 
function reads
\begin{equation}
Z=\int \mathrm{d}[\vec\phi]\, \e^{-H/T}\,,
\label{particion}
\end{equation}
the integration measure being the standard rotationally-invariant
measure on the sphere. In the following, expectation values
will be indicated as $\langle\dots\rangle$.

The zero temperature limit is dominated by the spin configurations
that minimize the energy. Exploratory Monte Carlo simulations showed
that these configurations have a bipartite structure. Indeed, let us
call a lattice site even or odd, according to the parity of the sum of
its coordinates ($\vec r_i=(x_i,y_i,z_i)$, $x_i+y_i+z_i$ even or
odd). Then the spins on the (say) even lattice are all parallel, while
the spins on the odd sublattice are randomly placed on a cone of angle
$\Omega$ ($\vec\phi_i\cdot \vec\phi_j=\cos
\Omega$) around the direction of the even lattice. The energy when $T$ goes 
to zero is simply
\begin{equation}
H_0(\Omega)= -3 L^3 \left(J \cos\Omega
+\sqrt{1+\cos\Omega}\right)\,.
\end{equation}
Now, $\Omega(J)$ is obtained by minimizing $H_0(\Omega)$.  For
$J>-\sqrt2/4$, $\Omega=0$, meaning a ferromagnetic vacuum. For
$-\sqrt2/4>J>-0.5$, one has $0<\Omega<\pi/2$ (ferrimagnetic vacuum),
while for all $J<-0.5$, it is $\pi/2<\Omega<\pi$ (antiferrimagnetic
vacuum). The full antiferromagnetic configuration ($\Omega=\pi$) is
never stable at zero temperature. The $J=-0.5$ point is very peculiar:
much like for the AFM-RP$^2$ model~\cite{SHROCK,ROMANO,RP2D3}, spins
in the even sublattice are randomly aligned or anti-aligned with the
(say) $Z$ axis, while the spins in the odd sublattice are randomly
placed in the $X,Y$ plane. Since spins in the two sublattices are
perpendicular to each other, one can arbitrarily reverse every spin. 
A local Z$_2$
symmetry is dynamically generated and, as we will see, it extends to
finite temperatures. An operational definition of dynamical generation
of a gauge symmetry is the following. One must calculate the
correlation-length for non gauge-invariant operators
($\xi_\mathrm{NGI}$) and compare it to the correlation-length
corresponding to gauge-invariant quantities ($\xi_\mathrm{GI}$). In
the continuum-limit ($\xi_\mathrm{GI}\to\infty$), one should have
$\xi_\mathrm{NGI}/\xi_\mathrm{GI}\to 0$. We have checked that the
correlation-length associated to the spin-spin correlation function
(non Z$_2$ gauge-invariant) is smaller than $0.3$ for all temperatures
at $J=-0.5$.
The alert reader will notice that the symmetry group at the
point $(T,J)=(0,-0.5)$ is rather a local O(2), besides the local Z$_2$
previously discussed. However, the associated correlation-length at
finite temperature grows enormously when approaching the critical
temperature (tenths of lattice-spacings already at $T=0.9
T_\mathrm{c}$), and probably diverges. More details on this will be
given in Ref.~\cite{CRUZ}.

A further analytical evidence for this fact can be obtained by performing a
Taylor expansion of the action at $J=-1/2$ assuming that
$\vec\phi_i\cdot \vec\phi_j$, which just vanishes at $J=-1/2$ and 
zero temperature, is small:
\begin{equation}
-H=1-{\textstyle \frac{1}{8}} \sum_{<i,j>}  (\vec\phi_i\cdot \vec\phi_j)^2 
+{\textstyle \frac{1}{16}} \sum_{<i,j>} (\vec\phi_i\cdot \vec\phi_j)^3
-{\textstyle \frac{5}{128}} \sum_{<i,j>} (\vec\phi_i\cdot \vec\phi_j)^4+\ldots
\end{equation}
We can assume that this expansion has a finite radius of convergence
and so we can extend this series to the non zero temperature region.
Notice that the first term in the expansion is just the AFM-RP$^2$
Hamiltonian modified by terms that are no longer gauge
invariant (those with odd powers in the scalar product). We can argue
that those terms, which break explicitly the  gauge invariance,
are irrelevant operators, in the Renormalization Group sense, at the
PM to AFM-RP$^2$ critical point and so, our model at finite temperature
should belong to the same Universality Class as the AFM-RP$^2$ model.
Obviously, were the transition of the first order, the argument would not
apply.

In the Z$_2$ gauge-invariant phase, the vectorial magnetizations defined
as
\begin{equation}
\begin{array}{lcl}
{\vec M}_\uu&=&\frac{1}{L^3}\sum_i  \vec\phi_i\,,\\
{\vec M}_\s&=&\frac{1}{L^3}\sum_i  (-1)^{x_i+y_i+z_i}\vec\phi_i\,,\\
\end{array}
\end{equation}
are zero. Thus, we define proper order parameters, invariant under that gauge
symmetry, in terms of the spin
field, $\vec\phi_i$ and the related spin-2 tensor field (which is
$Z_2$ invariant):
\begin{equation}
\mathbf{\tau}_i^{\alpha\beta}=\phi_i^\alpha\phi_i^\beta - 
\frac{1}{3}\delta^{\alpha\beta}\quad,\quad \alpha,\beta=1,2,3\,.
\end{equation}
Then we define:
\begin{equation}
\begin{array}{lcl}
\bbox{\tau}_i^\uu&=&\bbox{\tau}_i\,, \\
\bbox{\tau}_i^\s&=&(-1)^{x_i+y_i+z_i}\bbox{\tau}_i\,, \\
{\mathbf M}_{\uu,\s}&=&\frac{1}{V}\sum_i \bbox{\tau}_i^{\uu,\s}\,.
\end{array}
\end{equation}
The different phases we find (see~\cite{CRUZ} for details) are:
paramagnetic, ferromagnetic ($\langle{\vec
M}_\uu\rangle\neq 0$, $\langle{\vec
M}_\s\rangle= 0$), ferrimagnetic 
($\Vert\langle{\vec M}_\uu\rangle\Vert>\Vert\langle{\vec M}_\s\rangle\Vert>0$),
antiferrimagnetic ($\Vert\langle{\vec M}_\s\rangle\Vert>\Vert\langle{\vec
M}_\uu\rangle\Vert>0$), antiferromagnetic ($\langle{\vec
M}_\s\rangle\neq 0$, $\langle{\vec
M}_\uu\rangle= 0$), and RP$^2$
($\langle{\mathbf M}_\uu\rangle,\langle{\mathbf M}_\s\rangle\neq 0$ but with vanishing vectorial magnetizations). The phase
diagram can be seen in Fig.~\ref{FIGPHASEDIAG}. Notice the strong
similarities of the point $(T=0,J=-0.5)$ with a Quantum Critical
Point~\cite{QCP}. A detailed analysis of this phase diagram will
appear soon~\cite{CRUZ}. 

Let us end this section by defining the observables actually used in the
simulation. 
They are obtained in terms of the Fourier transform of
the tensor fields:
\begin{equation}
\widehat{\mathbf T}^{\uu,\s}_{ p}=\sum_{ \vec r \in L^3} 
\e^{-{\mathrm i}  \vec p\cdot \vec r}\ {\mathbf \tau}^{\uu,\s}_{\vec r},
\end{equation}
and their propagators
\begin{equation}
G_{\uu,\s}({\vec p})=\frac{1}{V}\left\langle{\mathrm{tr}} \widehat{\mathbf
T}^{\uu,\s}_{\vec p} \widehat{\mathbf T}^{\uu,\s\dag}_{\vec p} \right\rangle\,,
\label{PROPAGATORS}
\end{equation}
where
\begin{equation}
\vec p= \frac{2\pi}{L}\vec{n}\ ,\ n_i=0,\ldots,L-1\,.
\end{equation}
Notice that $G_\uu(0)=V {\mathrm{tr}} \langle ({\mathbf M}_\uu)^2\rangle$, and
$G_\uu(\pi,\pi,\pi)=G_\s(0,0,0)=V {\mathrm{tr}} \langle ({\mathbf M}_\s)^2\rangle$. Then
we have the usual ($\chi_\uu$) and staggered ($\chi_\s$) susceptibilities,
\begin{equation}
\chi_{\uu,\s}=G_{\uu,\s}(0,0,0)\,.
\end{equation}
Having those two order parameters, we must expect the following
behavior for the propagators in the thermodynamic limit, in the scaling
region and for $T>T_\cc$:
\begin{equation}
\begin{array}{rcl}
G^{\uu,\s}(\vec p)&\simeq&\displaystyle\frac{Z_{\uu,\s}\xi_{\uu,\s}^{-\eta_{\uu,\s}}}
{\vec p^2 +\xi_{\uu,\s}^{-2}} \,,\\ 
\xi_{\uu,\s}&=&A_{\uu,\s}t^{-\nu}(1+B_{\uu,\s}t^{-\nu\omega}+\ldots),
\end{array}\label{MYPROP}
\end{equation}
where $\xi_\uu$ and $\xi_\s$ are correlation-lengths,
$t=(T-T_\cc)/T_\cc$ is the reduced temperature and  $Z_{\uu,\s}$,
$A_{\uu,\s}$ and $B_{\uu,\s}$ are  
constants.  On the other hand, the anomalous
dimensions $\eta_\uu$ and $\eta_\s$ need not be equal: we can relate
them to the dimensions of the composite operators following the
standard way: $d-2+\eta_\uu = 2 \mathrm{dim}(\bbox{\tau}^\uu)$ and
$d-2+\eta_{\s} = 2 \mathrm{dim}(\bbox{\tau}^{\s})$, and in general 
$\mathrm{dim}(\bbox{\tau}^{\s}) \neq \mathrm{dim}(\bbox{\tau}^\uu)$.

To study the propagators (\ref{MYPROP}) on a finite lattice, we need 
to use the minimal momentum propagators
\begin{equation}
F_{\uu,\s}=\frac{1}{3}(G_{\uu,\s}(\pi/L,0,0)+ \mathrm{(permutations)})\,.
\end{equation}
With those quantities in hand, one can define a finite-lattice
correlation-length~\cite{COOPER} for the staggered, and non staggered
sectors:
\begin{equation}
\xi_{\uu,\s} =\left(\frac{\chi_{\uu,\s}/F_{\uu,\s}-1}
{4{\sin}^2(\pi/L)}\right)^{1/2}\,.\\
\end{equation}

Other quantities of interest are the dimensionless cumulants
\begin{equation}
\kappa_{\uu,\s}=
\frac{\langle\left({\mathrm{tr}}({\mathbf M}_{\uu,\s})^2\right)^2\rangle}
{\langle {\mathrm{tr}}({\mathbf M}_{\uu,\s})^2\rangle^2}\,.
\end{equation}
Besides the above quantities, we also measure the energy
(\ref{accion}), which is used in a reweighting
method~\cite{FERRSWEN}, and temperature derivatives of operators
through their crossed correlation with the energy.

\section{Critical behavior}

For an operator $O$ that diverges as $|t|^{-x_O}$,
its mean value at temperature $T$ in a size $L$
lattice can be written, in the critical region, assuming the finite-size
scaling ansatz as~\cite{LIBROFSS} 
\begin{equation}
O(L,T)=L^{x_O/\nu}\left(F_O(\xi(L,T)/L)+O(L^{-\omega})\right), 
\label{FSS}
\end{equation}
where $F_O$ is a smooth scaling function and $\omega$ is the
universal leading correction-to-scaling exponent.
 
In order to eliminate the unknown $F_O$ function, we use the method of
quotients~\cite{ON,ISPERC,RP2D3,PELISSE}. One studies the behavior of the operator
of interest in two lattice sizes, $L$ and $rL$ (typically $r=2$):
\begin{equation}
Q_O=O(rL,t)/O(L,t).
\end{equation}
Then one chooses a value of the reduced temperature $t$, such that the
correlation-length in units of the lattice size is the same in both
lattices~\cite{RP2D3}. One readily obtains
\begin{equation}
\left.Q_O\right|_{Q_{\xi}=r}=r^{x_O/\nu}+O(L^{-\omega}).
\label{QUO}
\end{equation}
Notice that the matching condition $Q_{\xi}=r$ can be easily tuned
with a reweighting method. The usual procedure consists on fixing
$r=2$, and obtaining the above quotients for several $L$ values
in order to perform an infinite volume extrapolation.

In order to obtain the critical exponents, we use as operators $\chi_{\uu,\s}$
($x_\chi=\gamma=2-\eta$), $\partial_T \xi_\s$ 
($x_{\partial_T \xi_\s}=\nu+1$). Notice that several
quantities can play the role of the correlation-length in Eq.~(\ref{QUO}):
$\xi_\uu,\xi_\s,L \kappa_\uu$ and $L\kappa_\s$. This simply changes the
amplitude of the scaling corrections, which will turn out to be quite
useful.

Another interesting quantity is the shift of the apparent critical
point ({\em i.e.}
$r\xi(L,T_\cc^{L,r})=\xi(rL,T_\cc^{L,r})$), with
respect to the real critical point:
\begin{equation}
T_\cc^{L,r}- T_\cc\propto
\frac{1-r^{-\omega}}{r^{\frac{1}{\nu}}-1}L^{-\omega-\frac{1}{\nu}}\,.
\label{SHIFTBETA}
\end{equation}

\section{The Simulation}

We have studied the model (\ref{accion}) in lattices
$L=6,8,12,16,24,32,48$ and $64$, with a Monte Carlo simulation
at $J=-0.5$. The
algorithm has been a standard Metropolis with 2 hits per
spin. The trial new spin is chosen randomly in the sphere. The
probability of finally changing the spin at least once is about $50\%$.

We have carried out 20 million full-lattice sweeps (measuring every 2 sweeps)
at each lattice size at $T=0.056$. For the $L=64$ lattice we have also 
performed 20 million sweeps at $T=0.0558$. 
The largest autocorrelation time measured is about 1400 sweeps (corresponding
to $\chi_\s$). To ensure the thermalization we have discarded a minimum
of 150 times the autocorrelation time. 

The computation was made on the RTN3 cluster of 28 1.9GHz PentiumIV processors
at the University
of Zaragoza and the total simulation time was equivalent to 11 months of a 
single processor.

\section{Critical exponents}

\begin{figure}[t]
\begin{center}
{\centering\epsfig{file=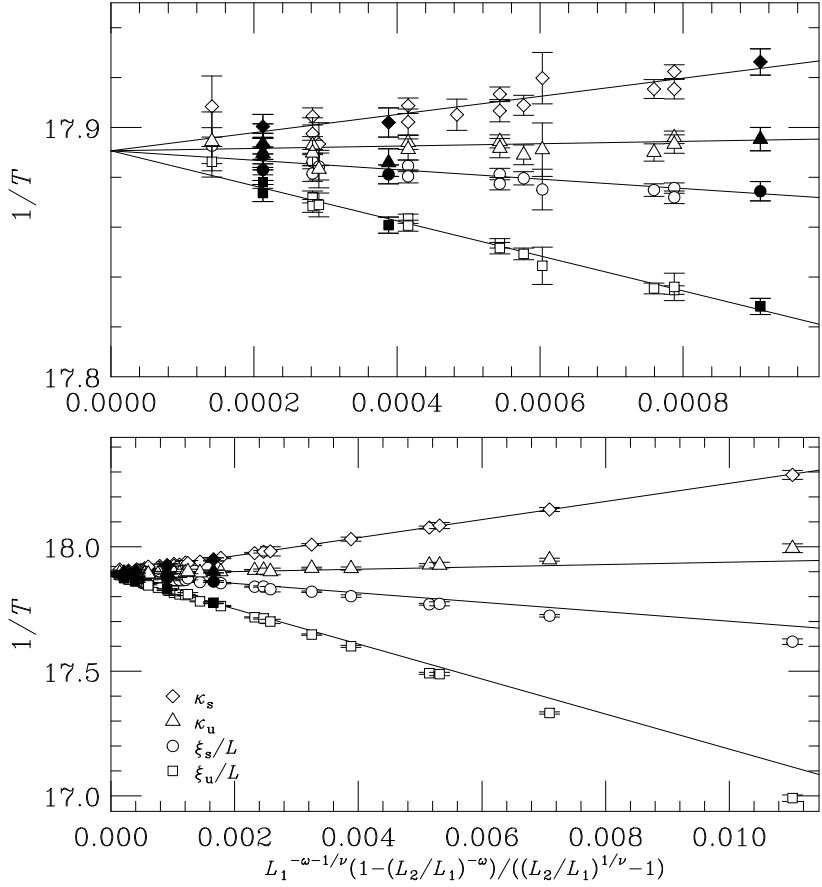,width=0.85\linewidth,angle=90}}
\end{center}
\protect\caption{Shift of the apparent critical temperature
Eq.~(\protect\ref{SHIFTBETA}) with the lattice size, using the four
possible kinds of dimensionless operators: 
$\xi_\uu/L$, $\xi_\s/L$, $\kappa_\uu$ 
and $\kappa_\s$. The top panel is a magnification of the leftmost region. 
Only filled data points, corresponding to $r=2$, were used in the fit.}
\label{FIGOMEGA}
\end{figure}

\begin{table}
\scriptsize
\begin{center}
\begin{tabular}{|cccccccc|}
\hline
$L_\mathrm{min}$&$T_\cc$&$\Delta T_\cc$&$\omega$&$\Delta\omega$&
$\chi^2/D$&$D$&$Q$\\
\hline
\hline
6 &0.0559075&0.0000034&0.959&0.021&3.11&23&0.00\\
8 &0.0558946&0.0000039&0.862&0.025&1.33&19&0.15\\
12&\bf0.0558951&\bf0.0000055&\bf0.817&\bf0.050&0.76&15&0.72\\
16&0.0558984&0.0000078&0.815&0.277&0.72&11&0.72\\
\hline
\end{tabular}
\end{center}
\caption{Results of the infinite volume extrapolation with
Eq.~(\ref{SHIFTBETA}) to obtain $T_\cc$ and $\omega$. $Q$ is the
quality-of-fit parameter. Our final values are the bold values.}
\label{TABLEOMEGA}
\end{table}

\begin{figure}[t]
\begin{center}
{\centering\epsfig{file=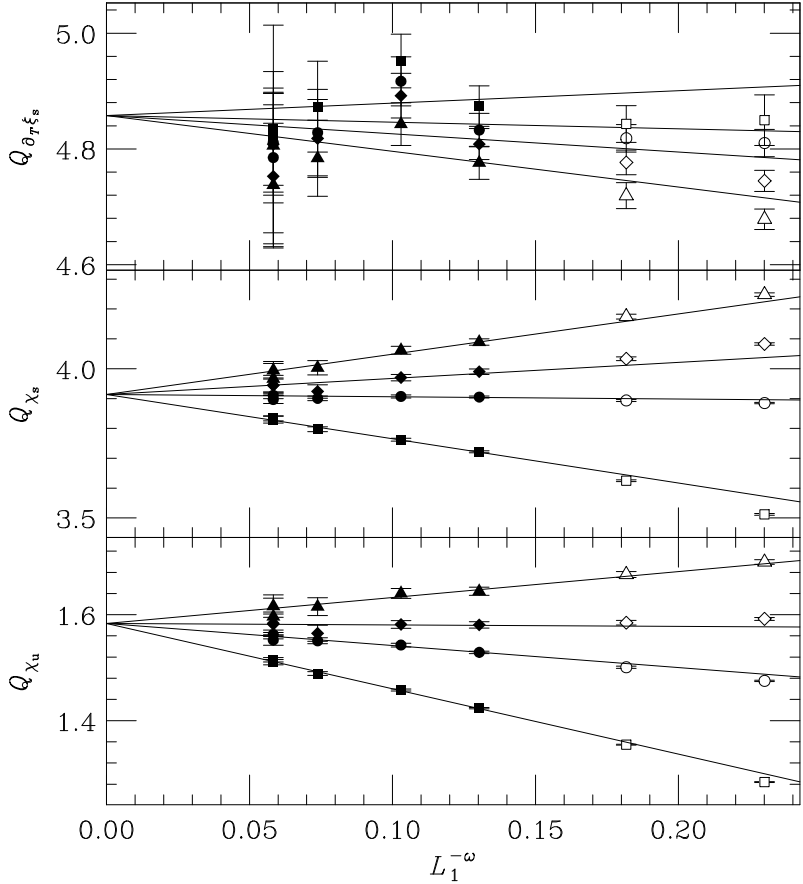,width=0.85\linewidth,angle=90}}
\end{center}
\protect\caption{Quotients defined in Eq.~(\protect\ref{QUO}), for
$\partial_T\xi_\s$ (top), $\chi_\s$ (medium) and $\chi_\uu$, as a function of
$L^{-\omega}$, for $r=2$. We have measured at the crossing of four 
dimensionless operators, $\xi_\uu/L$, $\xi_\s/L$, $\kappa_\uu$ and  
$\kappa_\s$. Symbols are as in Fig. 2\,.}
\label{FIGEXPONENTES}
\end{figure}

\begin{table}
\scriptsize
\begin{center}
\begin{tabular}{|cccccc|}
\hline
$L_\mathrm{min}$ & $\nu$&$\Delta \nu$&$\chi^2/D$&$D$&$Q$\\
\hline
\hline
6 &0.7722&0.0041&2.55&23&0.00\\
8 &0.7724&0.0055&1.66&19&0.03\\
12&\bf0.7811&0.0107&1.03&15&0.42\\
16&0.7898&\bf0.0179&0.83&11&0.61\\
\hline
&$\eta_\s$&$\Delta \eta_\s$&&&\\
\hline
 6&0.0238&0.0012&18.4&23&0.00\\
 8&0.0278&0.0016&1.97&19&0.01\\
12&\bf0.0315&0.0027&0.67&15&0.81\\
16&0.0359&\bf0.0049&0.56&11&0.87\\
\hline
&$\eta_\uu$&$\Delta \eta_\uu$&&&\\
\hline
6 &1.3259&0.0013&8.52&23&0.00\\
8 &1.3334&0.0018&1.56&19&0.06\\
12&\bf1.3368&0.0031&0.77&15&0.72\\
16&1.3435&\bf0.0058&0.62&11&0.82\\
\hline
\end{tabular}
\end{center}
\caption{Infinite volume extrapolation for the critical exponents, using 
Eq.~(\protect\ref{QUO}) and $\omega=0.82$. The fits were performed using
$L\geq L_\mathrm{min}$. The goodness-of-fit ($Q$) is also indicated. The
chosen extrapolation and error are emphasized.
}\label{TABLAEXPONENTES}
\end{table}

The first step, as usual, is to estimate $\omega$ from
Eq.~(\ref{SHIFTBETA}).  For this, one needs a rough-estimate of
$\nu$. Since our data for the quotient of $\partial_T \xi_\s$ using
$\xi_\s$ as correlation-length show very small scaling corrections (see
Fig.~\ref{FIGEXPONENTES}), one can temporally choose $\nu=0.78$ and
proceed with the determination of $\omega$.  Having four possible
dimensionless quantities, 
$\xi_\uu/L$, $\xi_\s/L$, $\kappa_\uu$ and $\kappa_\s$, we can
perform a joint fit to Eq.~(\ref{SHIFTBETA}) constrained to yield the
same $T_\cc$. This largely improves the accuracy of the final
estimate.  The full covariance matrix is used in the fit, and errors
are determined by the increase of $\chi^2$ by one-unit. Our results
are summarized in Fig.~\ref{FIGOMEGA} and table~\ref{TABLEOMEGA}.
Although scaling corrections are clearly visible, good fits are
obtained from $L_\mathrm{min}=12$. Thus, we conclude that
\begin{equation}
\omega=0.82(5)\ ,\ T_\cc=0.055895(5)\,.
\end{equation}
We are now ready for the infinite volume extrapolation of the critical
exponents. As usual, one needs to worry about higher-order scaling
corrections. Here we shall follow a conservative criterion: we shall
perform the fit to Eq.~(\ref{QUO}) only for $L\geq L_\mathrm{min}$, and
observe what happens varying $L_\mathrm{min}$. Once we found a
$L_\mathrm{min}$ for which the fit is acceptable and the infinite
volume extrapolation for $L\geq L_\mathrm{min}$ and $L>
L_\mathrm{min}$ are compatible within errors, we take the extrapolated
value from the $L\geq L_\mathrm{min}$ fit, and the error from the $L>
L_\mathrm{min}$ fit.  Our results can be found in
Fig.~\ref{FIGEXPONENTES} and in table~\ref{TABLAEXPONENTES}. As well
as for the critical temperature, we used all four dimensionless quantities
in a single fit constrained to yield a common infinite volume
extrapolation. Our final estimates are
\begin{equation}
\begin{array}{rclrclrcl}
\nu &=& 0.781(18)(1), &\eta_\s &=&0.032(5)(2),
&\eta_\uu&=&1.337(6)(7)\,,
\end{array}\label{NUFINAL}
\end{equation}
were the second error is due to the uncertainty in $\omega$.

One can compare these results with other models, once it is decided
what is going to play the role of our $\eta_\uu$ in the O($N$) models. Our
candidate is the tensorial representation~\cite{ON}\footnote{ In
O($N$) models it is possible to compute
$\eta_{\vec\phi\otimes\vec\phi}$ using field theoretical methods. We
should compute the dimensions of the operators $\phi_i ^2$ (scalar)
and $\phi_i \phi_j$ ($i \neq j$), for instance, introducing them in
correlation functions. The results are reported in \cite{Zinn} in
terms of the functions $\eta^{(1)}(g)$ (for the former operator) and
$\eta^{(2)}(g)$ (for the latter one). 
The anomalous dimension
of $\phi_i \phi_j$ ($i \neq j$) is $\eta_{\vec\phi\otimes\vec\phi}=d-2-2
\eta^{(2)}(g^*)$, where $g^*$ is the non trivial zero of the
$\beta$-function and $d=4-\epsilon$ is the dimension of the space.
The dimension of $\phi_i^2$ is just the dimension of
the energy operator: $\mathrm{dim}(\phi^2)=d-1/\nu$ and
$\eta_{\phi^2}=d-2-2 \eta^{(1)}(g^*)=d+2-2/\nu$. Only at the Mean
Field level $\eta_{\phi^2}=\eta_{\vec\phi\otimes\vec\phi}=2$ and $2
\beta_\phi = \beta_{\vec\phi\otimes\vec\phi}=1$.
Up to second order in  $\epsilon$ one has
$\eta_{\vec\phi\otimes\vec\phi}= 2-7 \epsilon /11 +133 \epsilon^2
/1331$ ($N\!=\!3$) and $\eta_{\vec\phi\otimes\vec\phi}= 2-2 \epsilon
/3 +\epsilon^2 /12$ ($N\!=\!4$). Setting $\epsilon=1$ one obtains
$\eta_{\vec\phi\otimes\vec\phi} \simeq 1.46$
($N\!=\!3$) and 1.42 ($N\!=\!4$), values not so far from the numerical
ones.}

\begin{equation}
\begin{array}{lrclcclccl}
\mathrm {RP}^2~\mbox{\protect\cite{RP2D3}}:&\nu &=& 0.783(11), &\eta_\s &=&0.038(3),
&\eta_\uu&=&1.339(10)\, ,\\
\mathrm{O}(3)~\mbox{\protect\cite{ON}}:&\nu &=& 0.704(6), &\eta_\phi &=&0.0413(16),
&\eta_{\vec\phi\otimes\vec\phi}&=&1.427(3)\, ,\\
\mathrm{O}(4)~\mbox{\protect\cite{ON}}:&\nu&=& 0.748(9), &\eta_\phi&=&0.0384(25),
&\eta_{\vec\phi\otimes\vec\phi}&=&1.374(5)\, .
\end{array}
\label{EXPOON}
\end{equation}

\section{Conclusions}
 
We have studied numerically a bosonic version of the DEM, Eq.~(\ref{accion}),
by Monte Carlo simulations,
obtaining its full phase-diagram, missed in previous
studies~\cite{PLAN1}.  We have studied its critical behavior with
Finite-Size Scaling techniques.
As Eq.~(\ref{NUFINAL}) and Eq.~(\ref{EXPOON}) show, our results for the
critical exponents are fully compatible with the results for the
AFM-RP$^2$ model, barely compatible with the O(4) model, and fully
incompatible with the results for the O(3) model. Our results in the
low temperature phase~\cite{CRUZ} seem to indicate that the scheme of
symmetry-breaking is O(3)/O(2), which contradicts Universality. Most
puzzling is the excellent agreement between the present results and
the estimates for the AFM-RP$^2$ model. This seems to indicate that
the AFM-RP$^2$ model really represents a new Universality Class in
three dimensions, in plain contradiction with the
Universality-Hypothesis (at least in its more general form). This
seems to imply that the local isomorphism of ${\cal G}/{\cal H}$ is
not enough to guarantee a common Universality Class. Of course, it
could happen that we were seeing only {\em effective} exponents and
that in the $L\to\infty$ limit a more standard picture arises. Yet, we
do not find any obvious reason for two very different models to have
such a similar {\em effective} exponents.

Another intriguing effect is that the augmented local $Z_2$ symmetry
of the point $(T=0,J=-0.5)$ extends to the finite temperature plane,
which is recalling a Quantum Critical Point~\cite{QCP}. Indeed, we
find that the Universality Class is the one of an explicitly
gauge-invariant model. To our knowledge this is a new effect in
(classical) Statistical Mechanics, and deserves to be called dynamical
generation of a gauge symmetry.

\section*{Acknowledgments}
We thank very particularly J.L. Alonso for pointing this problem to us,
for his encouragement and for many discussions.
It is also a pleasure to thank Francisco Guinea for discussions.  The
simulations were performed in the PentiumIV cluster RTN3 at the
Universidad de Zaragoza.  We thank the Spanish MCyT for financial
support through research contracts FPA2001-1813, 
FPA2000-0956, BFM2001-0718 and PB98-0842.
V.M.-M. is a Ram\'on y Cajal research fellow (MCyT).

\end{document}